\documentclass{article}

\usepackage[preprint]{neurips_2025}
\usepackage{graphicx}
\usepackage{amsmath}
\usepackage{multirow}

\usepackage[utf8]{inputenc} 
\usepackage[T1]{fontenc}    
\usepackage{hyperref}       
\usepackage{url}            
\usepackage{booktabs}       
\usepackage{amsfonts}       
\usepackage{nicefrac}       
\usepackage{microtype}      
\usepackage{xcolor}         

\title{Simple Models, Rich Representations: Visual Decoding from Primate Intracortical Neural Signals}

\author{Matteo Ciferri\thanks{These authors contributed equally to this work} \\
University of Rome, Tor Vergata\\
Department of Biomedicine and Prevention\\
\texttt{matteo.ciferri@students.uniroma2.eu} \\
\And
Matteo Ferrante\textsuperscript{*} \\
University of Rome, Tor Vergata\\
Department of Biomedicine and Prevention\\
\texttt{matteo.ferrante@uniroma2.it} \\
\And
Nicola Toschi \\
University of Rome, Tor Vergata\\
Department of Biomedicine and Prevention\\
A.A. Martinos Center for Biomedical Imaging \\
Harvard Medical School/MGH, Boston (US) \\
}

\begin{document}

\maketitle

\begin{abstract}
Understanding how neural activity gives rise to perception is a central challenge in neuroscience. We address the problem of decoding visual information from high-density intracortical recordings in primates, using the THINGS Ventral Stream Spiking Dataset. We systematically evaluate the effects of model architecture, training objectives, and data scaling on decoding performance. Results show that decoding accuracy is mainly driven by modeling temporal dynamics in neural signals, rather than architectural complexity. A simple model combining temporal attention with a shallow MLP achieves up to 70\% top-1 image retrieval accuracy, outperforming linear baselines as well as recurrent and convolutional approaches. Scaling analyses reveal predictable diminishing returns with increasing input dimensionality and dataset size. Building on these findings, we design a modular generative decoding pipeline that combines low-resolution latent reconstruction with semantically conditioned diffusion, generating plausible images from 200 ms of brain activity. This framework provides principles for brain-computer interfaces and semantic neural decoding.
\end{abstract}

\section{Introduction}

A complete account of perception and behavior must bridge neural representations with mental states, linking spikes and field potentials to the contents of subjective experience and overt action.  Recent progress in cognitive science and computational neuroscience has been catalyzed by three intertwined developments.  First, community-driven efforts now release large, meticulously curated datasets that pair rich sensory stimulation with high-resolution neural recordings \citep{NSDDataset,horikawa_generic_2017,bold5000,things}.  Second, advances in machine learning—particularly deep generative modeling and scalable optimization—provide expressive function classes capable of capturing the complex structure of brain-world mappings \citep{antonello2023scaling,oota_deep_2023,banville2025scalinglawsdecodingimages}.  Third, experimental practice is changing from \emph{wide} surveys of many individuals to \emph{deep}, longitudinal studies that expose a few subjects to tens of thousands of stimuli, drastically increasing statistical power \citep{Kupers2024}.

These factors have revived bidirectional modeling of the stimulus–brain relationship.  \emph{Encoding} models predict neural responses from sensory features, helping us understand the functional organization of the cortex, whereas \emph{decoding} models seek to reconstruct stimuli - or latent variables relevant to the task - from brain activity, a line of work central to basic science, as well as emerging brain–computer interfaces.  Successes span multiple modalities (EEG, MEG, fMRI, ECoG, and Utah array recordings) and cognitive domains, including language comprehension, speech production, music, and vision \citep{functional_review,oota_deep_2023}.  However, even with invasive data, key questions persist: What properties of intracortical spike trains carry the information necessary for high-fidelity decoding?  How do architectural choices—linear versus nonlinear models, temporal aggregation windows, loss functions—shape performance limits?  And how do these factors interact with scale, both in terms of training data and in terms of the dimensionality of neural input?

\begin{figure}[ht]
    \centering
    \includegraphics[width=0.95\linewidth]{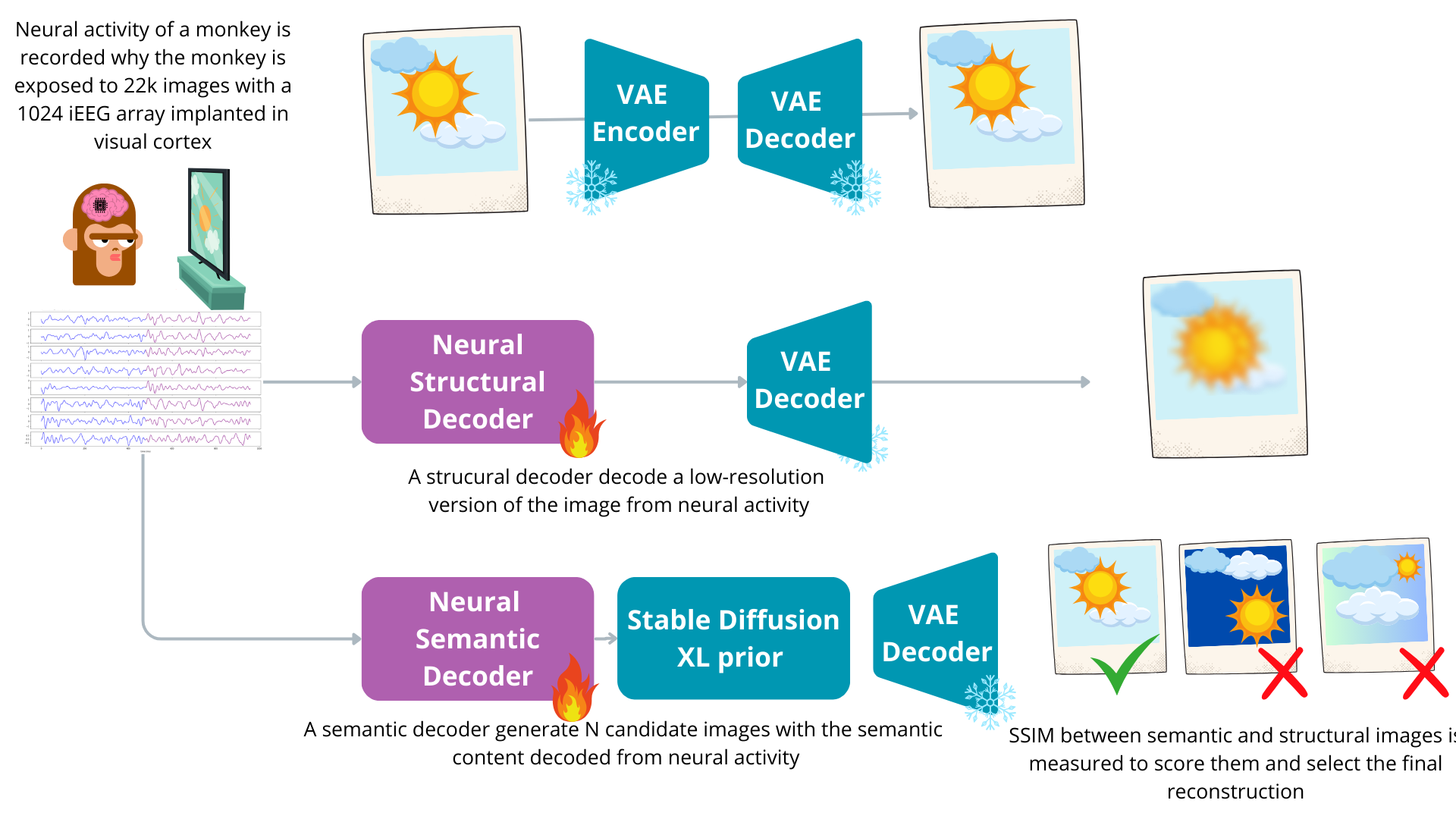}
    \caption{Overview of our generative decoding framework. Neural activity is recorded from a macaque implanted with a 1024-channel MUA array while viewing 22,000 natural images. A two-branch decoding pipeline is used: (top) a \textbf{structural decoder} maps neural signals to a low-resolution latent representation using a VAE decoder, providing a faithful but coarse reconstruction; (bottom) a \textbf{semantic decoder} generates multiple candidate images via Stable Diffusion XL conditioned on semantic information decoded from the same neural signal. The final image is selected by computing the SSIM between each candidate and the structural image, effectively combining the structural accuracy of the low-res decoder with the generative power of the semantic branch.}
    \label{fig:overview}
\end{figure}

We address these questions through the lens of visual decoding.  Our study leverages the recently released THINGS Ventral-Stream Spiking Dataset (TVSD) \citep{papale2025ventral}, in which two macaques viewed \(\sim25\,{\rm k}\) natural images drawn from 1,854 object categories while they recorded multi-unit activity (MUA) from \(\sim\!2{,}000\) channels distributed across V1, V4 and IT at 30 kHz.  Each image in the training partition was shown once, while 100 held-out images were repeated 30 times to boost signal-to-noise ratio and enable stringent cross-validation. 
In this work, we tackle the following research questions:
\begin{enumerate}
    \item \textbf{Temporal versus architectural complexity.}  
    What role does temporal structure play in neural decoding, and what types of models are best suited to capture it? While our experiments do not quantify the exact contribution of millisecond-scale timing, they suggest that modeling temporal dynamics is crucial for high-level decoding tasks. To explore this, we trained various models to decode neural data into semantic image representations obtained using a frozen CLIP model. We systematically compared architectures ranging from simple linear models to recurrent neural networks capable of capturing complex nonlinear dynamics. We show that the key driver of semantic decoding performance is the capacity to model the temporal structure in neural responses. Nonlinear models do improve performance, but our experiments reveal that their benefits are largely attributable to better temporal aggregation rather than to increased architectural complexity or spatial modeling. 
    The evaluation was carried out by measuring the accuracy of image retrieval on held out test data, specifically quantifying the model’s ability to identify exact images from neural activity using top-1 and top-5 retrieval metrics. 
    
    \item \textbf{Objective functions.}  We compare two ways of predicting vector representations from brain activity: mean squared error loss and representation alignment with contrastive learning. Similarly to the previous point, we used retrieval performance as a probe of the quality of decoded embeddings.
    
    \item \textbf{Scaling laws.}  We chart the performance of our best model in the retrieval task as a function of (i) the number of trials and (ii) the 
    number of principal components derived from neural channels, revealing predictable regimes of diminishing returns that inform experimental design.

    \item \textbf{Generative Decoding}  We introduce a two-stage decoder that samples candidate images from a frozen generative prior (Stable Diffusion \citep{podell2023sdxl}) and performs rejection sampling guided by a learned neural likelihood, achieving near-photorealistic reconstructions from \(\approx200\,\mathrm{ms}\) windows of activity. See Figure \ref{fig:overview} for an overview of our generative decoding pipeline.
\end{enumerate}

To address the first two research questions, we used zero-shot image retrieval as a proxy for assessing the quality of the mapping between brain activity and semantic visual representations. 
Crucially, we avoid generative models at this stage to minimize the confounding influence of strong image priors. When a generative model produces a high-quality image, it becomes difficult to discern whether the result reflects successful decoding or simply the model's inherent ability to produce photorealistic samples \citep{shirakawa2024spuriousreconstructionbrainactivity}. Retrieval-based evaluation offers a more transparent and interpretable benchmark: the model must identify the correct image from a fixed candidate pool based solely on the neural signal, allowing precise measurement of top-1 and top-5 accuracy.

However, retrieval has its limitations: mainly, its reliance on a predefined candidate set, which restricts generalization. For this reason, once we validated that our model achieved strong performance in this constrained setting, we turned to the more ambitious goal of generative decoding, confident that our brain to embedding map is robust due to prior validation in a retrieval setting. We propose a framework to extend brain-to-image mapping beyond fixed image sets, enabling open-ended visual reconstruction that overcomes the limitations of retrieval-only evaluation.


Together, these results introduce a practical decoding framework rather than a task-specific demo: (i) a lightweight temporal selection mechanism that turns noisy, high-rate intracortical activity into a stable semantic representation, and (ii) a modular retrieval-generation pipeline that separates “what was seen” (semantics) from “how it looked” (structure). This division makes the approach auditable, computationally efficient, and compatible with future closed-loop brain–computer interfaces.

\subsection{Related Work}\label{sec:related_work}

Recent years have seen great advances in decoding visual stimuli from neural activity, primarily in non-invasive settings such as fMRI \citep{oota_deep_2023,functional_review,antonello2023scaling,Gallant2012,huth2016natural,ferrante2023eyes,ferrante2024neuralfoundationmodelsvision,ferrante2024rbrhythmbrain}. Methods using pre-trained vision language models such as CLIP, paired with linear regression or contrastive learning, have enabled retrieval-based decoding and increasingly realistic image reconstruction when combined with diffusion models \citep{ozcelik2023braindiffuser, ferrante2023semantic, lin2022mind,scotti2023reconstructing,scotti2024mindeye2,chen2022seeing,xia2023dream}. In parallel, invasive modalities such as ECoG and multi-unit activity (MUA) have enabled higher-resolution decoding, though datasets of sufficient size and complexity have been scarce.
The release of the THINGS ventral stream spiking dataset (TVSD) \citep{papale2025ventral} has catalyzed progress by offering tens of thousands of diverse images paired with MUA from primate V1, V4 and IT, allowing detailed studies of visual representations with single-spike precision. The most closely related to our work is the MonkeySee study by \cite{MonkeySee}, which also uses TVSD and proposes a CNN-based end-to-end decoder with a learned inverse retinotopic mapping module to reconstruct images from neural signals. Their model emphasizes pixel-level realism and spatial interpretability, incorporating adversarial training and VGG feature losses to produce high-fidelity images.
In contrast, we focus on mapping between brain activity and semantic representations using zero-shot retrieval with frozen CLIP embeddings, sidestepping the confounding of strong generative priors. This setup allows us to systematically assess how model class (linear vs. nonlinear), temporal structure, and scaling of training set size impact decoding performance. Only after validating our brain-to-embedding map in this controlled setting, do we extend it to open-ended generative decoding using rejection sampling.
By investigating various research questions and proposing a separation between semantic decoding from image generation, our work distinguishes itself from MonkeySee’s direct reconstruction pipeline and offers deeper insight into the computational ingredients that enable robust visual decoding from invasive neural signals.

\section{Material \& Methods}

\begin{figure}[ht]
    \centering
    \includegraphics[width=.98\linewidth]{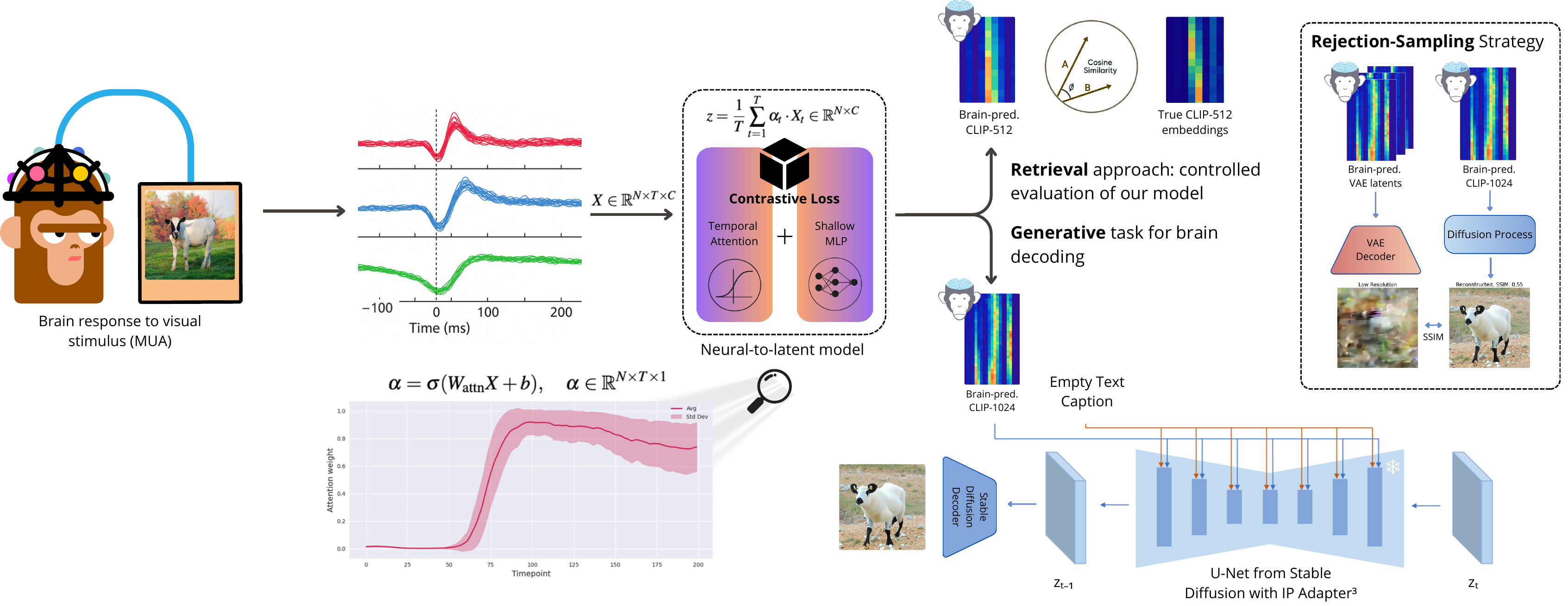}
    \caption{Overview of the proposed neural-to-semantic visual decoding pipeline. Multielectrode MUA responses evoked by visual stimuli are preprocessed and fed into a temporal-attention module that learns stimulus-dependent weighting over time, followed by a shallow MLP that maps neural activity into the CLIP embedding space. Training is supervised through a contrastive loss that aligns predicted neural embeddings with ground-truth CLIP representations of the viewed images. At test time, the predicted embeddings support two complementary tasks: (i) retrieval, where the model selects the most semantically similar image from a candidate set, and (ii) generation, where the embedding conditions a Stable Diffusion model (via an IP-Adapter) to synthesize novel images preserving the semantics inferred from neural activity. A rejection-sampling strategy ensures that generated samples remain structurally consistent with the original ones.}
    \label{fig:monkey_pipeline}
\end{figure}

We propose a brain decoding framework to estimate the embedding of a visual stimulus directly from neural recordings (Figure \ref{fig:monkey_pipeline}). The goal is to reconstruct a meaningful representation of the perceived image from intracortical signals, allowing two downstream applications: (i) \textit{ stimulation retrieval}, where the estimated embedding is compared against a set of candidate stimuli, and (ii) \textit{image generation}, where the predicted embedding is used as input to a generative model to synthesize a recognizable version of the original visual input.

\subsection{Data}

We conduct our analysis on both \textbf{Monkey F} and \textbf{Monkey N}, from the THINGS Ventral Stream Spiking Dataset (TVSD) \citep{papale2025ventral}. This dataset comprises intracortical multi-unit activity (MUA) recorded from 15 Utah arrays implanted in visual areas V1, V4, and IT of two macaque monkeys. Neural responses were collected while the monkeys passively viewed 22,248 unique natural images from the THINGS database, covering a broad distribution of object categories.
Each image was presented once for 200 ms, interleaved with 200 ms of a gray screen. The recordings were sampled at 30 kHz and temporally aligned with stimulus onset. For decoding, we used a 200 ms post-stimulus window. A subset of 100 images was 
presented 30 times each to the same subject and held out for evaluation, while 10\% of the training set samples were held out as the validation set for hyperparameter optimization. This design enables both training on a large-scale dataset and reliable, low-noise testing.
The neural data were preprocessed as follows. Let \( X_{\text{raw}} \in \mathbb{R}^{N \times T \times C} \) denote the raw MUA data, with \( N \) the number of samples, \( T = 200 \) timepoints, and \( C = 1024 \) channels. The neural data were standardized by z-scoring each channel across all timepoints and samples in the training set. The same normalization parameters were then applied to the test set.

\subsection{Neural Model}

In this section, we describe our proposed decoding model. The target representations \( Y \in \mathbb{R}^{N \times D} \), with \( D = 512 \), consist of high-level visual embeddings corresponding to the presented images, computed from the pretrained CLIP visual encoder~\citep{clip}.

In order to learn a mapping from MUA signals to the corresponding image embedding, we propose a neural architecture that can take into account the time evolution of the neural response. The model is designed to attend over the temporal dimension of the neural sequence and project the aggregated representation to the 512-dimensional target space.

Given an input tensor \( X \in \mathbb{R}^{N \times T \times C} \), the model computes a soft attention over time points:
\[
\alpha = \sigma(W_{\text{attn}} X + b), \quad \alpha \in \mathbb{R}^{N \times T \times 1}
\]
where \( W_{\text{attn}} \) denotes a linear layer and \( \sigma \) is the activation of the sigmoid. The attended representation is computed as:
\[
z = \frac{1}{T} \sum_{t=1}^{T} \alpha_t \cdot X_t \in \mathbb{R}^{N \times C}
\]
The vector \( z \) is then projected into the output space via a multilayer perceptron consisting of two fully connected layers with GELU activation and dropout:
\[
\hat{Y} = W_2 \cdot \text{Dropout}(\text{GELU}(W_1 z + b_1)) + b_2, \quad \hat{Y} \in \mathbb{R}^{N \times D}
\]

The training objective is a \textit{contrastive loss} based on cosine similarity between predicted and ground-truth embeddings. Let \( S \in \mathbb{R}^{N \times N} \) be the cosine similarity matrix between \( \hat{Y} \) predicted outputs and \( Y \) targets:
$
S_{ij} = \frac{\hat{y}_i^\top y_j}{\|\hat{y}_i\| \|y_j\|}.
$
The loss function is a variant of the NT-Xent loss with temperature scaling \( \tau \), learned during training:
\[
\mathcal{L}_{\text{contrastive}} = - \frac{1}{N} \sum_{i=1}^{N} \log \left( \frac{\exp(S_{ii} / \tau)}{\sum_{j=1}^{N} \exp(S_{ij} / \tau)} \right)
\]

In order to contextualize the performance of our proposed model, we compare it against several standard baselines commonly used in brain decoding literature. The following baseline models were considered:

\begin{itemize}
    \item Linear Model with Temporal Attention: similar to the proposed model, this variant uses the same temporal attention mechanism, but replaces the final MLP with a single linear layer to project the representation to the CLIP embedding space.
    \item Linear Model with Temporal Averaging: a linear regression trained on neural features obtained by averaging the neural signal over the temporal dimension, i.e., reducing each trial from \( \mathbb{R}^{T \times C} \) to \( \mathbb{R}^{C} \).
    \item Linear Model on Flattened Input: a linear regression trained on the fully flattened MUA signal, reshaped from \( \mathbb{R}^{T \times C} \) to a 1D-vector \( \mathbb{R}^{T \cdot C} \).
    \item MLP with Temporal Averaging: a feedforward neural network trained on the same time-averaged representation as above, introducing non-linearity over the input features at channel-wise level.
    \item Recurrent neural network: an LSTM processes the neural sequence over time. The last hidden state of the LSTM is extracted and passed through a projection layer to obtain the predicted embedding.
    \item Temporal Convolutional Network: a model composed of stacked 1D convolutional layers, followed by adaptive average pooling to reduce the temporal dimension. The pooled representation is then passed through an MLP to produce the final embedding. 
\end{itemize}

Hyperparameters, including learning rate, batch size, network depth for deep models, and regularization strength, were selected based on the performance of the validation set. The models were trained with both contrastive learning and standard regression objectives. All experiments were carried out on a high performance server equipped with eight NVIDIA A100 GPUs (80 GB each, interconnected via NVLINK), 256 CPU threads and 2 TB of system memory.

\subsection{Task 1: Stimulus Retrieval}

In order to assess the quality of the predicted embeddings, we performed a retrieval task, where each predicted embedding \( \hat{y}_i \) from the test set is matched against all ground truth embeddings \( \{y_j\}_{j=1}^{N} \), and the nearest neighbors are retrieved based on cosine distance.
After generating predictions for all test samples in evaluation mode, we collected both the predicted embeddings \( \hat{Y} \in \mathbb{R}^{N \times D} \) and the corresponding ground truth embeddings \( Y \in \mathbb{R}^{N \times D} \). The cosine distance was used to compute nearest neighbors in the embedding space:
$
\text{dist}(\hat{y}_i, y_j) = 1 - S_{ij} = 1 - \frac{\hat{y}_i^\top y_j}{\|\hat{y}_i\| \|y_j\|}.
$
For each test sample \( i \), we identified the top-\( k \) most similar ground truth embeddings over the test set. We computed the proportion of test samples for which the nearest neighbor is the ground truth target embedding, i.e., when the index of the closest neighbor matches the sample index (Top-1 Accuracy). We also evaluated the proportion of test samples for which the ground truth embedding appears within the top-5 retrieved neighbors (denoted as Top-5 Accuracy). 
This retrieval setup provides a quantitative measure of the semantic similarity between predicted and target embeddings, and acts as an indirect proxy of brain to image representation mapping quality in form of a decoding metric.

\subsection{Scaling Laws}

In order to investigate how the performance of the decoding model scales with different properties of the input data, we conducted two different sets of controlled experiments evaluating the impact of (i) the input dimensionality, and (ii) the size of the training set (i.e., number of available samples). We applied Principal Component Analysis (PCA) to the MUA signals across channels to analyze input dimensionality impact on performance. We fitted PCA models with different components, and projected both the training and test sets into the reduced channel subspace. The resulting PCA-reduced data had shape \( \mathbb{R}^{N \times T \times C'} \), where \( C' < C \) is the selected number of components. This allowed us to test the model under different values of \( C' \) while keeping the temporal resolution constant.
To evaluate the effect of training data size on model performance, we subsampled the training set in a different scenario using a random selection of \( N' \) samples, with \( N' < N \). Specifically, we randomly selected a fixed number of training samples, using a controlled random seed for reproducibility. The test set remained fixed across all experimental conditions.
For each experimental condition—defined by a particular number of PCA components or training samples—we trained a new instance of the neural model from scratch and evaluated it on the full test set. This procedure enabled us to characterize the scaling behavior of the model under varying data availability and spatial resolution, highlighting how model performance is affected by data constraints commonly encountered in neural decoding settings.

\subsection{Task 2: Image Generation}

In addition to retrieval-based evaluation, we explore a generative setting in which the predicted neural embeddings are used to synthesize images. The goal of this task is to demonstrate that the decoded representations retain sufficient visual semantics to condition a generative model and reconstruct a plausible version of the original stimulus.
We use the Stable Diffusion model as the generative backbone \citep{podell2023sdxl}. In order to condition the model on high-level visual embeddings, we incorporate the IP-Adapter module~\citep{ye2023ip} into the diffusion pipeline. This adapter enables conditioning via learned visual representations rather than textual prompts. We first train our neural model with output dimensionality matching that of the IP-Adapter input (e.g., 1280) to predict visual embeddings from MUA data. In a second setup, we train the same model to directly predict the flattened latent representation of the image expected by the Stable Diffusion VAE (i.e., a tensor of shape \( 4 \times 32 \times 32 \)). Given the predicted visual embedding and structural latents, we generate different images for each test sample using classifier-free guidance. The predicted structural latents are only used to evaluate the low-resolution version of the image estimated from neural activity. 
Our approach is inspired by the recent trend of increasing test-time computation in AI systems~\citep{damian2022self, ji2025test}. Leveraging the strong mapping between brain activity and semantic content, and the ability of generative models to reconstruct semantically coherent images, we first generate $N$ candidate semantic images for each trial. Simultaneously, we evaluate the low-resolution image reconstruction that preserves the primary structure (i.e., overall shape and color) from brain activity. We further compare the generated images with the low-resolution preview decoded by the VAE using the predicted latents, and computing structural similarity (SSIM)~\citep{wang2004image} to rank the outputs. The reconstructed images with the highest SSIM scores are reported in the "Results" section.

\section{Results}

\subsection{Retrieval Accuracy}

We evaluated the decoding performance using a retrieval task in which the predicted visual embeddings were matched against the ground truth embeddings of all test samples. The model was considered successful if the correct target appeared within the top-\( k \) nearest neighbors, with Top-1 and Top-5 accuracy used as metrics.
To qualitatively assess the decoding performance, we select some random images and visualize the retrieved nearest neighbors in image space based on the predicted CLIP embeddings with our model. For each test image, we display the original stimulus alongside its top-5 neighbors retrieved from the test set (See  Supplementary Materials).
Table~\ref{tab:retrieval_accuracy} reports the retrieval accuracy across different decoding models and feature processing strategies. Our best-performing model combines temporal attention with a shallow MLP, achieving the highest retrieval performance and outperforming both linear baselines and more complex models such as LSTMs.

\begin{table}
\centering
\caption{Retrieval performance averaged over five seeds and two primates with different decoding models (using all channels). Best results in bold.}
\vspace{0.6em}
\label{tab:retrieval_accuracy}
\begin{tabular}{|l|cc|cc|}
\hline
\multirow{2}{*}{Decoding Model} & \multicolumn{2}{c|}{Top-1 Accuracy} & \multicolumn{2}{c|}{Top-5 Accuracy} \\
\cline{2-5}
& \rule{0pt}{2ex} MSE Loss & NT-Xent Loss & MSE Loss & NT-Xent Loss \\
\hline
Linear/TimeFlat     & 10.1\% $\pm$ 2.10\% & 41.3\% $\pm$ 3.92\% & 27.3\% $\pm$ 2.89\% & 72.6\% $\pm$ 5.66\% \\
Linear/AvgTime      & 21.4\% $\pm$ 1.36\% & 54.9\% $\pm$ 1.53\% & 45.0\% $\pm$ 2.97\% & 86.0\% $\pm$ 1.50\% \\
MLP/AvgTime         & 24.0\% $\pm$ 1.10\% & 66.1\% $\pm$ 2.08\% & 50.6\% $\pm$ 2.42\% & 90.1\% $\pm$ 2.48\% \\
LSTM                & 11.0\% $\pm$ 1.41\% & 37.7\% $\pm$ 2.06\% & 30.2\% $\pm$ 1.60\% & 73.7\% $\pm$ 2.37\% \\
TCN                 & 17.0\% $\pm$ 2.39\% & 58.3\% $\pm$ 3.11\% & 44.1\% $\pm$ 2.56\% & 86.6\% $\pm$ 1.77\% \\
Linear/TimeAtt      & 19.4\% $\pm$ 3.11\% & 62.7\% $\pm$ 2.79\% & 42.8\% $\pm$ 2.70\% & 89.4\% $\pm$ 1.34\% \\
MLP/TimeAtt         & 22.4\% $\pm$ 3.14\% & \textbf{69.3\%} $\pm$ \textbf{2.38\%} & 47.8\% $\pm$ 3.54\% & \textbf{93.6\%} $\pm$ \textbf{1.03\%} \\
\hline
\end{tabular}
\end{table}

\subsection{Scaling Laws}

\begin{figure*}[ht]
    \centering
    \includegraphics[width=.95\linewidth]{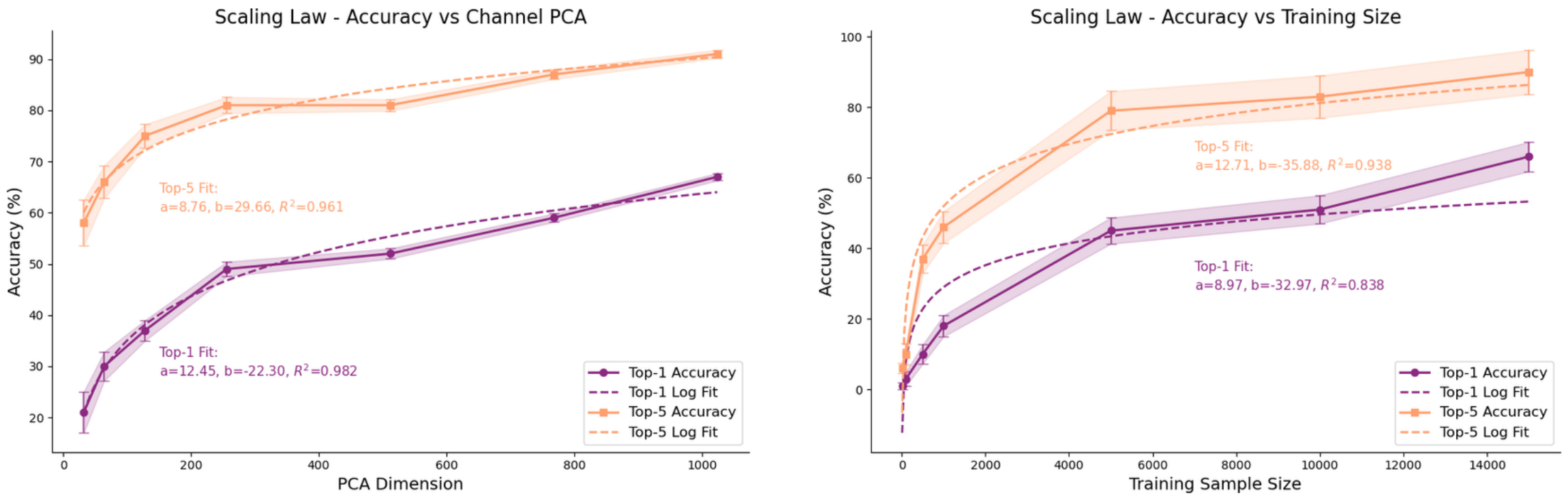}
    \caption{Left: Top-1 and Top-5 classification accuracy as a function of PCA dimensionality applied to neural channels. Accuracy increases logarithmically with dimensionality, as shown by high R$^2$ in the log-fit curves. 
    Right: Top-1 and Top-5 accuracy as a function of training set size. Performance scales log-linearly with data, underscoring the importance of dataset size in brain-based visual decoding.}
    \label{fig:scaling_laws}
\end{figure*}

We investigated how model performance scales with neural feature dimensionality and the size of the training dataset. As illustrated in Figure \ref{fig:scaling_laws} (left), increasing the number of principal components retained after applying PCA to the neural data consistently improves top-1 and top-5 classification accuracy. The trends are approximately logarithmic, as confirmed by a log-fit model, with large initial gains followed by diminishing returns beyond 256 dimensions. This supports the idea that while semantic information spans a high-dimensional neural space, a low-rank subspace can capture most of the information required for coarse-grained decoding.
On the right (Fig \ref{fig:scaling_laws}), we observe a strong scaling trend with respect to training set size. Performance improves rapidly as the number of training examples increases, particularly between 100 and 5000 samples. Although gains persist beyond 10,000 samples, they begin to taper off, indicating a regime of diminishing returns. Both plots exhibit classic scaling law behavior \citep{sato2024scalinglawneuraldata,antonello2023scaling,banville2025scalinglawsdecodingimages,kaplan2020scalinglawsneurallanguage}, where more data or higher-capacity representations improve performance predictably, albeit with sublinear returns.

\subsection{Attention Weights}

\begin{figure}
    \centering
    \includegraphics[width=.92\linewidth]{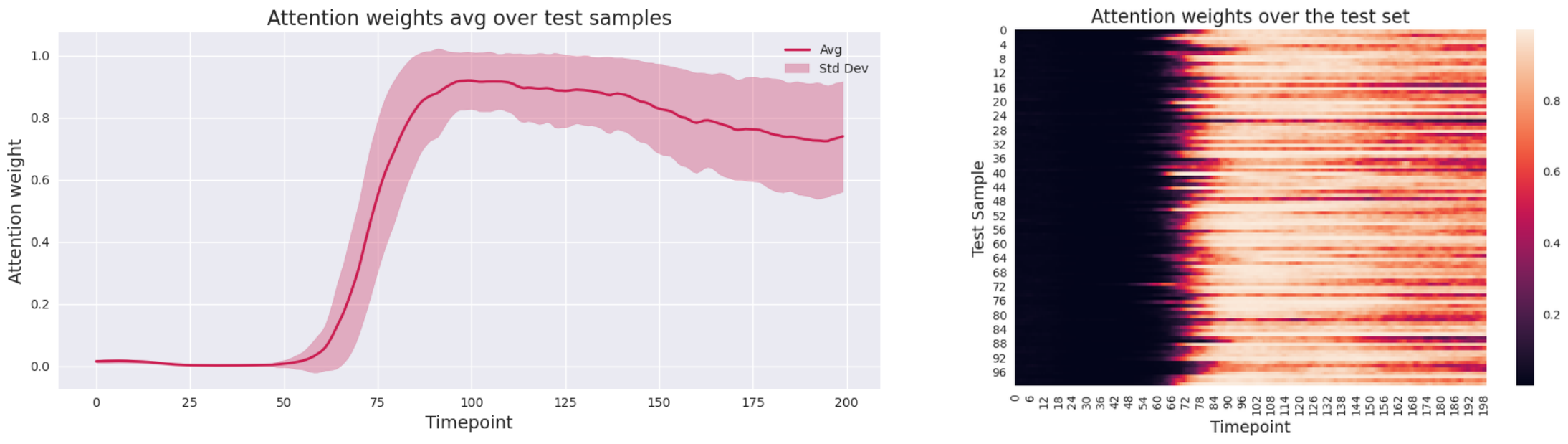}
    \caption{Heatmap of attention weights for the test set (right side) and average weights plot over the entire set (left side). Warmer colors of the heatmap indicate higher attention weights.}
    \label{fig:attention_results}
\end{figure}

In order to gain insight into the temporal focus of the decoding model, we extracted and visualized the attention weights produced by our neural model over the entire test set. For each input sample \( x \in \mathbb{R}^{T \times C} \), the model outputs a sequence of attention scores \( \alpha \in \mathbb{R}^{T} \) reflecting the relative importance of each timepoint in the final prediction.
During inference, we aggregated the attention weights for all test samples and constructed a matrix \( A \in \mathbb{R}^{N \times T} \), where each row corresponds to a test trial and each column to a timepoint. The attention analysis reveals interpretable patterns in the model’s temporal sensitivity (See Fig \ref{fig:attention_results}). The lightweight temporal selection module assigns data-driven weights to each millisecond of multi-unit activity, and this selective temporal integration is more important for accurate semantic decoding than increasing architectural depth or recurrence.

\subsection{Image Reconstruction}

\begin{figure}[ht]
    \centering
    \includegraphics[width=.95\linewidth]{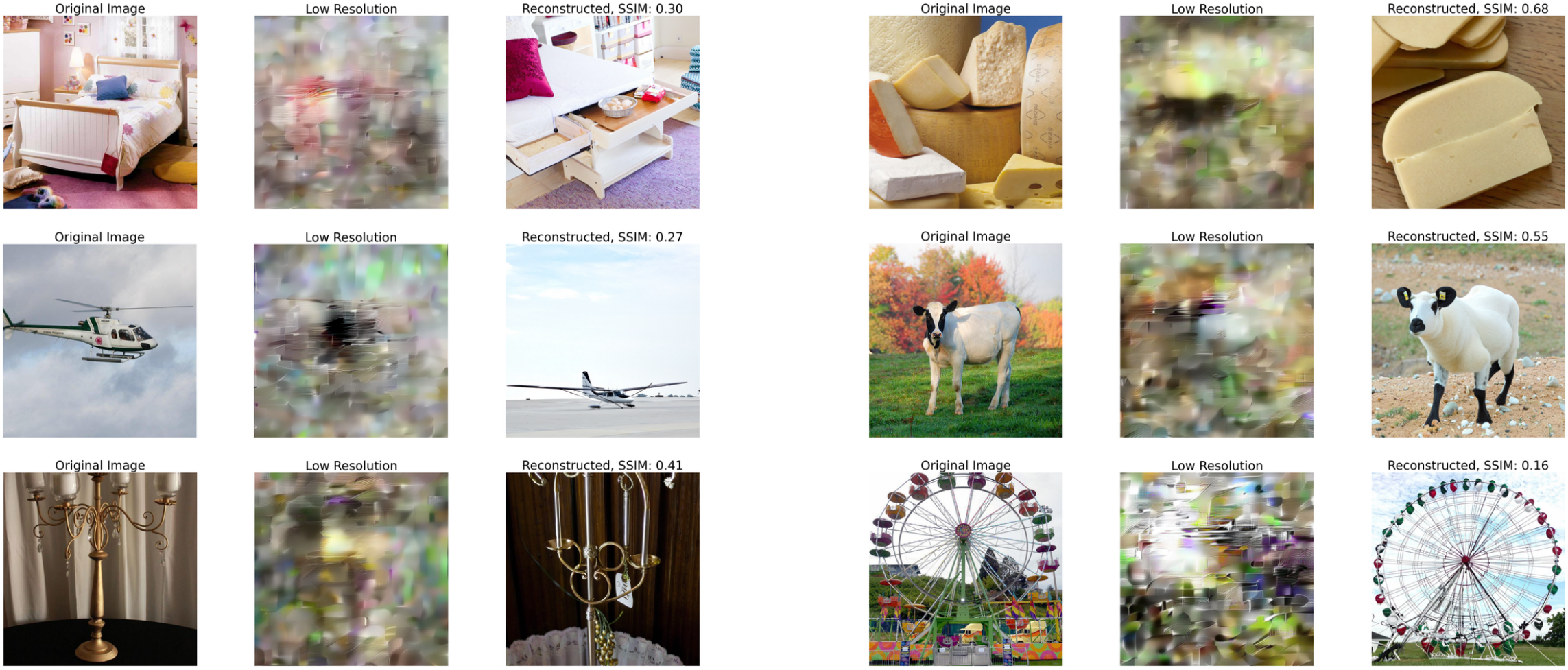}
    \caption{Examples of image reconstructions from neural activity. Each triplet shows the original image (left), a low-resolution baseline (middle), and our reconstruction (right) with corresponding SSIM score. The generations capture essential attributes such as object structure, color distribution, and general content.}
    \label{fig:gener_monkey}
\end{figure}

We evaluated the generative potential of our brain decoding model by estimating the latent representations required to guide the pretrained diffusion model. We adopt a semantic-first rejection sampling strategy, in which multiple candidate images are generated from the predicted embedding and ranked according to their structural consistency with a low-resolution preview (using the VAE component of the model) decoded from brain activity. 

Figure~\ref{fig:gener_monkey} shows representative examples of the image generation results. For each test sample, we display the original stimulus, the brain-inferred low-resolution preview, and the high-resolution reconstruction generated by the diffusion model conditioned on the estimated visual embedding. Reconstructions are ranked on the basis of the structural similarity index (SSIM).

\section{Discussion}

We investigated which factors influence visual decoding accuracy from primate intracortical recordings the most. We focused on temporal dynamics, model complexity, loss functions, and data scaling. This work proposes a practical recipe for brain-to-image decoding under realistic constraints. First, we show that a compact temporal selection module can turn noisy, prone-to-overfitting neural activity into a stable semantic representation that directly supports zero-shot identification of the viewed stimulus. Second, we turn that semantic representation into plausible reconstructions using a modular generation pipeline that explicitly separates semantic identity from structural layout via rejection sampling. In the following, we summarize our key findings and their broader implications.

\textbf{Non-linearity matters---mostly for temporal modeling}: To address our first research question, we show that gains from non-linearity are most pronounced when applied to models that preserve temporal resolution, rather than to temporally averaged inputs. Our soft-attention temporal aggregator, combined with a shallow MLP, consistently outperformed both linear baselines and recurrent architectures. The observed outcomes support our core claim that selective temporal integration matters more than modeling complex temporal dependencies end-to-end. High-resolution neural data contain many low-informative timepoints; sequence models propagate all of them, which increases noise sensitivity. In contrast, our soft-attention mechanism learns to upweight the most informative intervals, effectively filtering out noisy or irrelevant fluctuations.

\textbf{Retrieval-based evaluation improves interpretability}: Throughout our analyses, we initially prioritized retrieval-based evaluation over direct image generation as a metric of decoding success. By mapping neural activity to a fixed, semantically grounded embedding space (CLIP), we bypassed the confounding influence of strong generative priors that often obscure whether high-quality outputs really reflect accurate brain decoding \citep{shirakawa2024spuriousreconstructionbrainactivity}. 

\textbf{Scaling laws reveal predictable tradeoffs}:
Investigating the scaling behavior revealed consistent trends \citep{antonello2023scaling,banville2025scalinglawsdecodingimages}. Increasing the input dimensionality 
by retaining more principal components from the PCA decomposition led to substantial improvements in decoding accuracy, but with diminishing returns beyond approximately 256 components, suggesting that a relatively low-dimensional neural manifold suffices for capturing most semantic information. Similarly, scaling the number of training trials sharply improved performance, with top-5 retrieval accuracy increasing up to 90\% as the training set approached 15,000 examples.
Together, these scaling laws provide practical guidance for experimental design: to optimize decoding, the most effective thing is to acquire more diverse trials. Increasing input dimensionality—such as through higher channel count—also contributes to performance gains. Next-generation BCI systems could prioritize large-scale data collection and ultra-high-density recordings to further enhance decoding accuracy.

\textbf{Contrastive loss vs MSE loss}:
Across multiple decoding architectures, NT-Xent-trained models consistently achieve higher Top-1 and Top-5 accuracy. Beyond empirical performance, the outcome is also grounded in theoretical work: recent findings \citep{whyconcepts2024} argue that conceptual and semantic representations are best modeled as directional vectors in high-dimensional spaces. In this view, the direction of a vector encodes most of the meaningful structure, making cosine similarity—sensitive to direction but not to scale—particularly well-suited for tasks involving semantic embeddings such as ours.

\textbf{Generative decoding benefits from modularity}:
Building on robust retrieval performance, we developed a two-stage generative decoding pipeline. Rather than directly mapping neural data to high-resolution images, we split the task into semantic candidate generation via a frozen diffusion model and low-resolution structure reconstruction via direct latent-space prediction. Rejection sampling based on structural similarity (SSIM) allowed us to combine the best elements of both streams: semantic accuracy and structural faithfulness. This modularity avoided entangling the training objective with generative biases and offers a flexible blueprint for future decoding systems capable of both recognition and synthesis.

\textbf{Complementarity with existing approaches}:
Compared to contemporaneous efforts such as MonkeySee \citep{MonkeySee}, which focus on pixel-level reconstruction using CNN-based spatial mappings, our work emphasizes semantic decoding, temporal modeling, and principled evaluation through retrieval metrics. Together, these approaches are complementary: high-fidelity spatial reconstructions and semantically aligned representations represent two sides of the brain decoding challenge. Bridging them could offer a richer, multi-level understanding of neural information flow.


\textbf{Broader impact}:
Our study contributes to the development of decoding systems with potential relevance for human brain–computer interfaces (BCIs). By combining high-density recordings with a large number of trials, we observed substantial improvements in decoding performance. 
Although current results are limited to few subjects and single modality, the modular and interpretable nature of our pipeline—ranging from linear and attention based decoders to generative models—provides a foundation for future extensions across individuals, species, and recording techniques.

\subsection{Limitations}
Despite these promising results, several limitations warrant discussion. First, our retrieval-based evaluation is inherently constrained by the fixed candidate pool of images, which may underestimate the full expressive power of the decoded representations. Second, our semantic mappings rely on frozen CLIP embeddings, which were optimized for human visual-textual associations and may not perfectly align with primate visual representations \citep{xu2023imagereward}. Third, although our rejection-sampling strategy mitigates biases, the use of pretrained generative models still introduces priors that could inflate perceived reconstruction quality. Fourth, we cannot rule out the possibility that more complex, nonlinear architectures might achieve higher decoding performance. Finally, the passive viewing task minimized behavioral confounds but precluded the study of top-down modulations such as attention or memory, which could significantly shape neural dynamics during naturalistic behavior. Together, these limitations suggest fruitful directions for future research aiming to develop more robust, flexible, and biologically grounded decoding systems. Looking ahead, applications in humans raise critical concerns about neural privacy and consent \citep{yuste2017four}, calling for ethical considerations alongside technical progress.

\subsection{Conclusions}

Our findings show that lightweight, temporally aware models can accurately decode semantics from invasive primate recordings, with temporal dynamics being key to performance. A modular retrieval-generative framework supports flexible and interpretable reconstructions, paving the way for future extensions to cross-subject generalization, complex behaviors, and multisensory data.

\section*{Data and Code Availability}

All data are from a publicly available dataset (THINGS Ventral Stream Spiking Dataset) at: \url{https://gin.g-node.org/paolo_papale/TVSD}. Implementation code for reproducibility is available at the repository: \url{https://github.com/fidelioc55/primates-mua-decode}.

\bibliography{neurips_2025}
\bibliographystyle{abbrv}

\appendix

\section{Appendix}
This section provides additional qualitative and quantitative results to support the main findings of the article, particularly with regard to the advantages of contrastive learning in the brain decoding task. 

\begin{figure}[ht]
    \includegraphics[width=.95\linewidth]{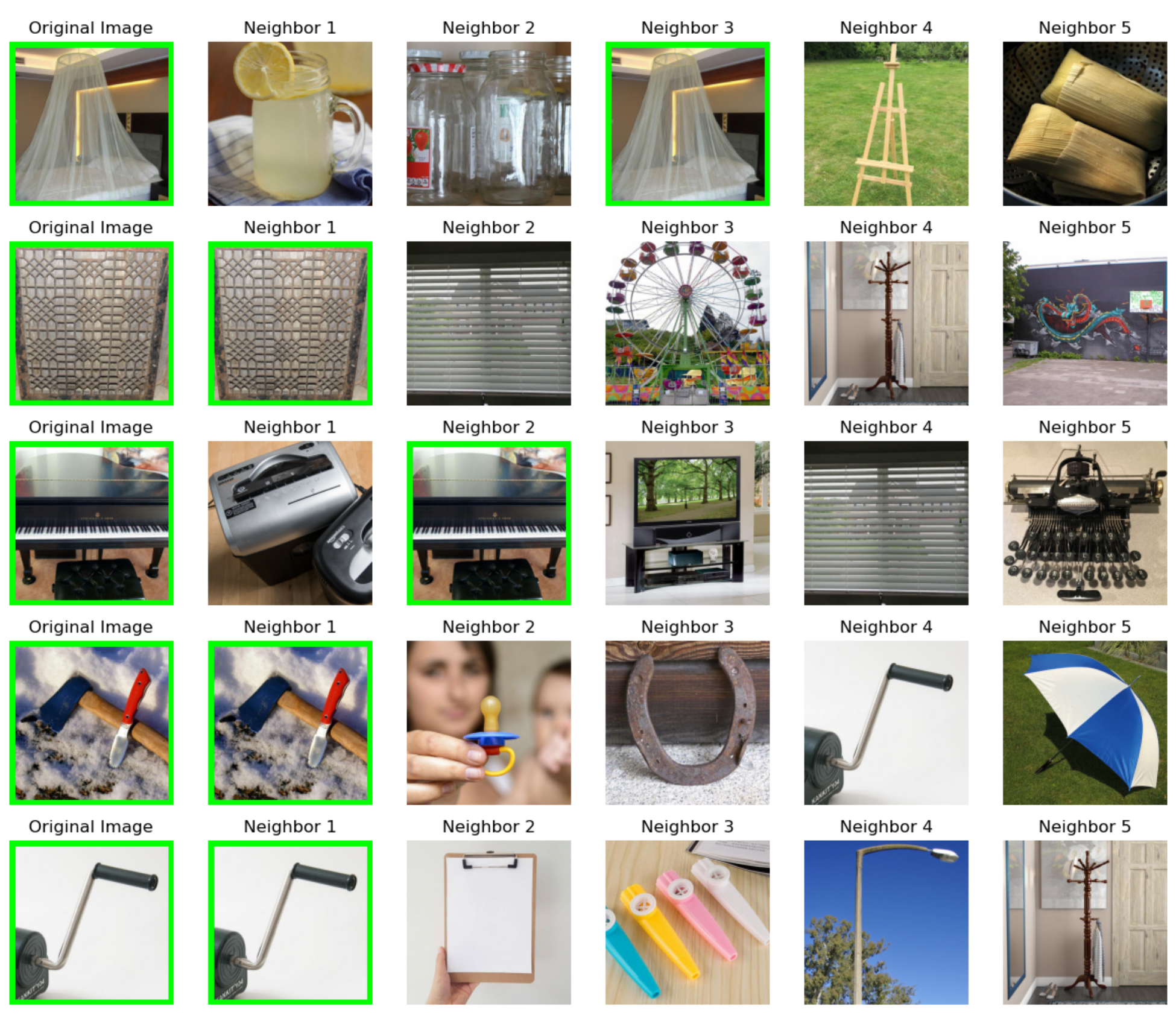}
    \caption{Top-5 image retrieval examples based on predicted embeddings. Each row shows one test sample: the original image (left) and the five nearest neighbors retrieved from the test set.}
    \label{retrieve_monkey_2}
\end{figure}

Figure \ref{retrieve_monkey_2} illustrates the qualitative retrieval results obtained using our proposed model trained with contrastive loss, discussed in the main article.

\subsection*{Evaluation of Decoding Model}

Table \ref{decoder_metrics_2} reports the evaluation metrics in image reconstruction for the best three model configurations. Across all metrics, the three variants exhibit similar performance, with subtle differences depending on the evaluation criterion. MLP with time attention shows better results in perceptual and semantic metrics (e.g., CLIP), indicating stronger high-level feature alignment. Linear model or averaging the time information performs well on loss-like metrics and pixel correlation, suggesting effective representation matching and favoring low-level reconstruction.
These results underscore the importance of evaluating reconstruction not just with pixel-wise losses but also with perceptual and embedding-based metrics, which better reflect the semantic fidelity of the decoded stimuli.

\begin{table}
\centering
\small
\begin{tabular}{lccc}
\toprule
Metric & MLP/TimeAtt & Linear/TimeAtt & MLP/AvgTime \\
\midrule
Pixel Correlation $\uparrow$     & 0.140 ± 0.163 & 0.151 ± 0.167 & \textbf{0.156 ± 0.163} \\
SSIM $\uparrow$                  & \textbf{0.364 ± 0.199} & 0.356 ± 0.202 & 0.342 ± 0.206 \\
MSE $\downarrow$                   & 0.110 ± 0.020 & 0.111 ± 0.024 & \textbf{0.107 ± 0.018} \\
Cosine Similarity $\uparrow$     & 0.811 ± 0.108 & \textbf{0.815 ± 0.123} & 0.801 ± 0.100 \\
InceptionV3 $\uparrow$           & \textbf{0.868 ± 0.225} & 0.836 ± 0.236 & 0.808 ± 0.230 \\
CLIP $\uparrow$                  & \textbf{0.879 ± 0.201} & 0.842 ± 0.229 & 0.827 ± 0.245 \\
EffNet Distance $\downarrow$       & \textbf{0.792 ± 0.144} & 0.822 ± 0.139 & 0.827 ± 0.137 \\
SwAV Distance $\downarrow$         & \textbf{0.492 ± 0.111} & 0.516 ± 0.114 & 0.528 ± 0.117 \\
\bottomrule
\end{tabular}
\vspace{1.0em}
\caption{Quantitative evaluation of three decoder variants across multiple metrics. Metrics marked with $\uparrow$ are better when higher (e.g., similarity or structural alignment), while those with $\downarrow$ are better when lower (e.g., error or distance). Bold values indicate the best score per row.}
\label{decoder_metrics_2}
\end{table}

Table \ref{feature_correlation_2} compares feature correlations across AlexNet layers between our MLP/TimeAtt model (generative and retrieval frameworks) and the MonkeySee baselines. Focusing on the generative variant (third column), our model consistently outperforms both MonkeySee Spatial and Spatiotemporal in the early convolutional layers (conv1 and conv2). This suggests that the rejection sampling procedure, which selects generated images based on low-level structural similarity from a pool of candidates, effectively enhances alignment with early visual features encoded in the brain. In the deeper fully connected layers (FC6–FC8), our model maintains strong performance, surpassing MonkeySee. This indicates that the generative decoder is also capable of capturing high-level semantic content, highlighting the model’s ability to represent both perceptual structure and abstract semantics from neural activity.

\begin{table}
\centering
\small
\begin{tabular}{lcccc}
\toprule
\textbf{Layer} & \textbf{MonkeySee Spatial} & \textbf{MonkeySee ST} & \textbf{MLP/TimeAtt (Gen)} & \textbf{MLP/TimeAtt (Retr)} \\
\midrule
conv1   & 0.358 & 0.372 & 0.528 & 0.874 \\
conv2   & 0.320 & 0.334 & 0.368 & 0.819 \\
conv3   & 0.429 & 0.443 & 0.335 & 0.808 \\
conv4   & 0.385 & 0.401 & 0.319 & 0.804 \\
conv5   & 0.292 & 0.318 & 0.305 & 0.798 \\
FC6     & 0.344 & 0.377 & 0.522 & 0.839 \\
FC7     & 0.534 & 0.579 & 0.500 & 0.827 \\
FC8     & 0.579 & 0.610 & 0.712 & 0.884 \\
\bottomrule
\end{tabular}
\vspace{1.0em}
\caption{Feature correlation (mean Pearson) between AlexNet features extracted from reconstructed and original images, across different decoding models (MonkeySee as a baseline).}
\label{feature_correlation_2}
\end{table}

\subsection*{Temporal Information}
We quantitatively assess how our model use temporal structure by comparing it to a millisecond level baseline. Specifically, we implemented a sliding estimator, training a separate MLP for each timepoint (millisecond resolution). Each MLP maps the neural signal at a single timepoint to the target embedding and is evaluated independently. This simulates what decoding would look like at high temporal precision (millisecond-scale), without aggregating across time. Our main model instead learns to combine all 200 timepoints with a self-attention mechanism, assigning learned weights to each.

\begin{figure}
    \centering
    \includegraphics[width=.85\linewidth]{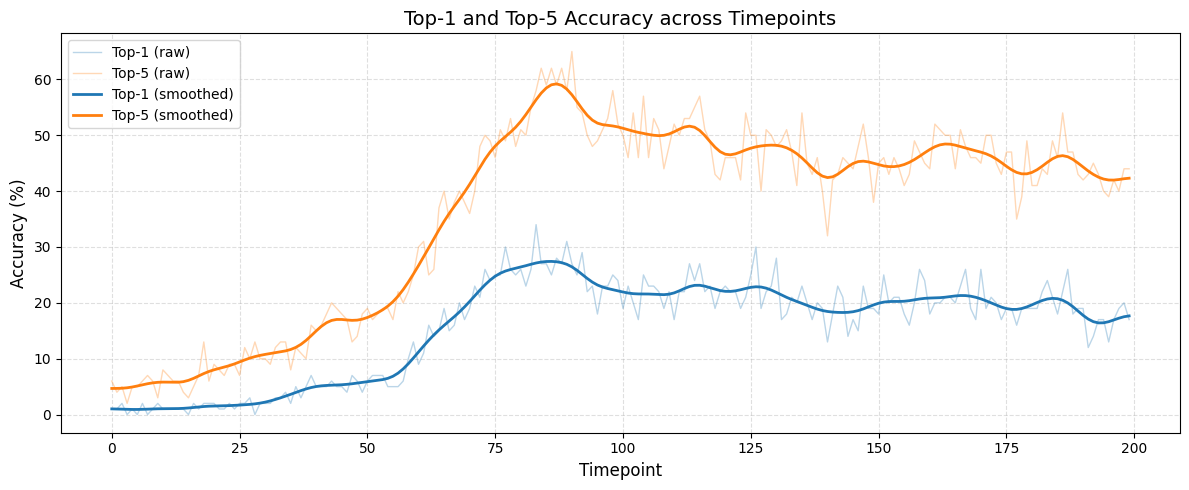}
    \caption{Decoding performance with MLP trained on a sliding estimator at millisecond level. The overall shape of performance is mirrored by our model but shifted on the y axis towards much worse performances.}
    \label{tmp_mlp_2}
\end{figure}

As shown in the Figure \ref{tmp_mlp_2}, our soft-attention model consistently outperforms the sliding baseline at each timepoints (best performance: around 30\% Top1, and 60\% Top5), and the curve of accuracy wrt timepoints follows the same behaviour of our temporal attention. This demonstrates that not just high-resolution data matters, but the ability of the model to aggregate and weigh temporal information in a data-driven way.

\subsection*{Hyperparameter Tuning}

We performed an hyperparameter search for each model architecture to identify the optimal configuration, as reported in Table \ref{best_hyperparams_2}. Across all models, the best results were consistently obtained using a contrastive loss.

\begin{table}
\centering
\renewcommand{\arraystretch}{0.90}
\begin{tabular}{@{}llc@{}}
\toprule
\textbf{Model} & \textbf{Hyperparameter} & \textbf{Best Value} \\
\midrule
\multirow{6}{*}{TCN} 
& Learning Rate     & 1e-3  \\
& Conv. Channels    & 256   \\
& Hidden Dim.       & 256   \\
& Kernel Size       & 7     \\
& Num. Layers       & 2     \\
& Loss Type         & CL    \\
\midrule
\multirow{5}{*}{LSTM} 
& Learning Rate     & 1e-4  \\
& LSTM Hidden Dim.  & 512   \\
& MLP Hidden Dim.   & 512   \\
& Num. Layers       & 2     \\
& Loss Type         & CL    \\
\midrule
\multirow{2}{*}{Linear (TimeFlat)} 
& Learning Rate     & 1e-3  \\
& Loss Type         & CL    \\
\midrule
\multirow{2}{*}{Linear (AvgTime)} 
& Learning Rate     & 1e-3  \\
& Loss Type         & CL    \\
\midrule
\multirow{4}{*}{MLP (AvgTime)} 
& Learning Rate     & 1e-3  \\
& Num. Layers       & 2     \\
& Hidden Dim.       & 768   \\
& Loss Type         & CL    \\
\midrule
\multirow{4}{*}{MLP (TimeAtt)} 
& Learning Rate     & 1e-3  \\
& Num. Layers       & 2     \\
& Hidden Dim.       & 768   \\
& Loss Type         & CL    \\
\bottomrule
\end{tabular}
\vspace{0.5em}
\caption{Best hyperparameter configuration per model.}
\label{best_hyperparams_2}
\end{table}

\end{document}